\documentclass[preprint,prc,aps,showpacs,groupedaddress,floatfix]{revtex4}
\usepackage{epsfig}
\usepackage{graphicx}

\begin{document}
\title{Global Optical Potential for the Elastic Scattering of $^{6}$He at Low Energies}
\author{Y. Kucuk}
\affiliation{Department of Physics, Giresun University, Giresun,
Turkey}
\author{I. Boztosun}
\affiliation{Department of Physics, Akdeniz University, Antalya,
Turkey,}
\author{T. Topel}
\affiliation{Institute of Science, Erciyes University, Kayseri,
Turkey}
\begin{abstract}

A set of global optical potential has been derived to describe the
interactions of  $^{6}$He  at low energies. The elastic scattering
angular distribution data measured so far for many systems, ranging
from $^{12}$C to $^{209}$Bi, have been considered within the
framework of the optical model in order to find a global potential
set to describe the experimental data consistently. We report that
very good agreement between theoretical and experimental results has
been obtained with small $\chi^{2}/N$ values by using the derived
potential set. The reaction cross section and volume integrals of
the potentials have been deduced from the theoretical calculations
for all studied systems at relevant energies.
\end{abstract}

\maketitle
\section{Introduction}
Defining the structure and dynamics of the halo nuclei has been a
central area for the nuclear physics  in the past decades.
Particularly, nuclear astrophysicists has been involved with the
reaction mechanism of the short-lived exotic nuclei, which bear
great importance due to the capture reactions that occurred in early
universe. To get more information regarding the nature of halo
nuclei and its reaction mechanism, many experiments have been
carried out by using the Radioactive Ion Beams (RIB) facilities.In
this respect, $^{6}$He has been one of the most studied nuclei to
understand the structure of the weak binding and of the large radial
extent to investigate the effect of the halo structure on the
reaction observables
\cite{Alkhalili,Alkhalili1,Karataglidis2000,Suzuki1991,Varga1994,Funada1994,Navratil2005,Sanchez-Benitez2005,Agu1,Agu2,Kolata,Kakuee2006,Kakuee2003,Raab1,Borowska2007,Korsheninnikov1997,Gasgues,Warner}.
In these works, the elastic scattering, the fusion and the
break-up/transfer cross sections  have been measured and studied
theoretically for many systems  at energies near the Coulomb barrier
to investigate the behavior of the optical potential and the effect
of break-up coupling to the reaction and the scattering mechanism.
The role of the Coulomb and nuclear break-up on the fusion cross
section has been attempted to be addressed by studying the
interaction of $^{6}$He  with heavy nuclei such as  $^{208}$Pb,
$^{209}$Bi and $^{238}$U
\cite{Agu1,Agu2,Kolata,Young,Trotta,Raab,Kolata1,Keeley,Keeley1,Keeley2,Matsumoto,Canto1,Canto,Sanchez}.
Different interpretations have been presented about how the break-up
coupling affects the fusion process. These works have been extended
from the heavy nuclei to weaker ones such as $^{27}$Al, $^{64}$Zn
and $^{65}$Cu and it has been observed that transfer and break-up
cross section were more important than the fusion cross sections at
energies above the Coulomb barrier for weaker systems  and total
cross section of the reactions induced by halo nuclei has a large
value as compared the total cross section of stable nuclei reactions
such as $^{4}$He and $^{6}$Li \cite{Benjamim,Pietro,Navin}.

In addition to discussions about the reaction mechanism of the halo
nuclei, the explanation of the measured elastic scattering angular
distributions  near the Coulomb barrier has been the other
motivation of these studies since elastic scattering bears  a great
importance to provide an idea about the nuclear optical potential of
the system. To observe the scattering mechanism of $^{6}$He, the
experimental data for many systems including light or heavy nuclei
have been analyzed by using  phenomenological and microscopic
potentials \cite{Smith,Milin}. In a recent paper, Milin \textit{et
al.}  \cite{Milin} have studied  $^{6}$He+$^{12}$C system and they
have measured the elastic and inelastic scattering as well as $2n$
transfer reaction angular distributions at $E_{Lab}$=18.0~MeV.They
have analyzed these data by using the Woods-Saxon shaped
phenomenological optical potential \cite{Milin} and they were able
to obtain a consistent agreement for the elastic scattering and
transfer reaction data, but they were not able to obtain the
inelastic 2$^+$ data simultaneously with the elastic and transfer
channels data. The same data has been analyzed by Boztosun
\textit{et al.}  \cite{Boztosun} and they were able to obtain a
simultaneous description of the elastic, inelastic and transfer
reaction cross sections by deforming the long range imaginary
potential within the framework of the CCBA formalism.

Studies on the elastic scattering of $^{6}$He on medium mass target
nuclei has been presented by some authors in previous years.
Benjamim \textit{et al.} \cite{Benjamim} have measured the elastic
scattering angular distribution of the $^{6}$He+$^{27}$Al system
with RIBRAS facility and have investigated the behavior of the total
reaction cross section. They have used the S$\tilde{a}$o Paula
Potential (SPP) to reproduce the elastic scattering data and they
have extracted the reaction cross section for this system at some
energies. For $^{6}$He+$^{64}$Zn system, elastic scattering angular
distributions, transfer/break-up angular distributions and fusion
excitation functions have been measured at near the Coulomb barrier
energies by Pietro \textit{et al.}\cite{Pietro} to investigate the
effects of neutron halo structure on the reaction mechanism. An
optical model analysis has been performed to explain the elastic
scattering data and the total reaction cross section data has been
extracted from this analysis. Another reaction of $^{6}$He on medium
mass target is  $^{6}$He+$^{65}$Cu system. For this system, the
measured elastic scattering cross section has been analyzed by using
the statistical model and the reaction cross section has been
obtained from the theoretical results \cite{Navin}.

$^{6}$He+$^{208}$Pb and $^{6}$He+$^{209}$Bi are the examples of the
systems with heavy targets, which have been studied extensively to
measure elastic scattering around the Coulomb barrier energies. The
$^{6}$He+$^{208}$Pb system has been recently  studied  by
S\'{a}nchez Ben\'{\i}tez \cite{Sanchez} and they have measured
elastic scattering cross section at energies between 14 and 22 MeV.
In this work, the experimental data have been analyzed by using the
phenomenological Wood-saxon potential and the presence of the long
range absorption has been reported for this system. Aguilera
\textit{et al.} \cite{Agu1,Agu2} for $^{6}$He+$^{209}$Bi system have
performed the simultaneous analysis of the elastic scattering and
transfer reaction cross section at energies below the Coulomb
barrier by using the optical model.

As seen from the literature, $^{6}$He interactions at energies
around Coulomb barrier have crucial importance in understanding the
properties of exotic systems and a global potential set is required
in the theoretical analysis of the reactions. So far, many potential
sets have been used either phenomenological or of the folding type
to describe the elastic scattering and other scattering observables
of $^{6}$He nucleus. These potentials are very similar to those of
the $^{6}$Li potentials. Sometimes, $^{4}$He potential has also been
used by adjusting the radius for $^{6}$He one.Although a good
description of the observables by using these potentials has been
obtained for individual reactions, there is no global potential that
describe the elastic scattering of $^{6}$He from different target
nuclei consistently.

Therefore, in this paper, we aim to develop a global potential set
to describe the elastic scattering of the $^{6}$He nucleus from
light to heavy target nuclei at low energies.  In the next section,
we present the optical model and introduce our global potential.The
results of the theoretical analysis by using our global potential
set for many systems have been presented in Section \ref{results}.
We conclude in Section \ref{conc}.

\section{OPTICAL MODEL CALCULATIONS}
We have performed an extensive study for the elastic scattering of
$^{6}$He on different target, from $^{12}$C to $^{209}$Bi, for a
wide energy range. We have used the optical model for the
theoretical calculations and the total effective potential in the
optical model consists of the Coulomb, centrifugal and nuclear
potentials as
\begin{equation}
V_{total}(r)= V_{Nuclear}(r)+ V_{Coulomb}(r)+ V_{Centrifugal}(r)
\label {Vtotal}
\end{equation}

In the total effective potential, the Coulomb and Centrifugal
potentials are well-known. The Coulomb potential \cite{Satchler} due
to a charge $Z_{P}e$ interacting with a charge $Z_{T}e$ distributed
uniformly over a sphere of radius $R_{c}$ is given by
\begin{eqnarray}
V_{Coulomb}(r) & = & \frac{1}{4\pi\epsilon_{\circ}}\frac{Z_{P}Z_{T}e^{2}}{r}, \hspace*{3cm} r\geq R_{c} \\
     & = & \frac{1}{4\pi\epsilon_{\circ}}\frac{Z_{P}Z_{T}e^{2}}{2R_{c}}(3-
\frac{r^{2}}{R_{c}^{2}}), \hspace*{1.3cm} r < R_{c} \label{coulomb}
\end{eqnarray}
where $R_{c}$ is the Coulomb radius, taken as 1.2 fm in the
calculations and $Z_{P}$ and $Z_{T}$ denote the charges of the
projectile $P$ and the target nuclei $T$ respectively.

The centrifugal potential is
\begin{equation}
V_{Centrifugal}(r) = \frac{\hbar^{2} \, l \, (l+1)}{2 \, {\mu} \,
r^{2}}
\end{equation}
where $\mu$ is the reduced mass of the colliding pair.

Finally, the complex $V_{Nuclear}(r)$ potential is taken to be the
sum of the Woods-saxon square shaped real and Wood-Saxon shaped
imaginary potentials given as

\begin{equation}
V_{nuclear}(r) =
\frac{-V_{0}}{\left[1+e^{\frac{r-R_{V}}{a_{V}}}\right]^{2}} +i
\frac{-W_{0}}{1+e^{\frac{r-R_{W}}{a_{W}}}} \label{nuclear}
\end{equation}

Here, $R_{i}$=$r_{i} [A_{P}^{1/3}+A_{T}^{1/3}]$ ($i=V$ or $W$),
where $A_{P}$ and $A_{T}$ are the masses of projectile and target
nuclei and $r_{V}$ and $r_{W}$ are the radius parameters of the real
and imaginary parts of the nuclear potential respectively.

\begin{figure}
\epsfxsize 16.0cm \centerline{\epsfbox{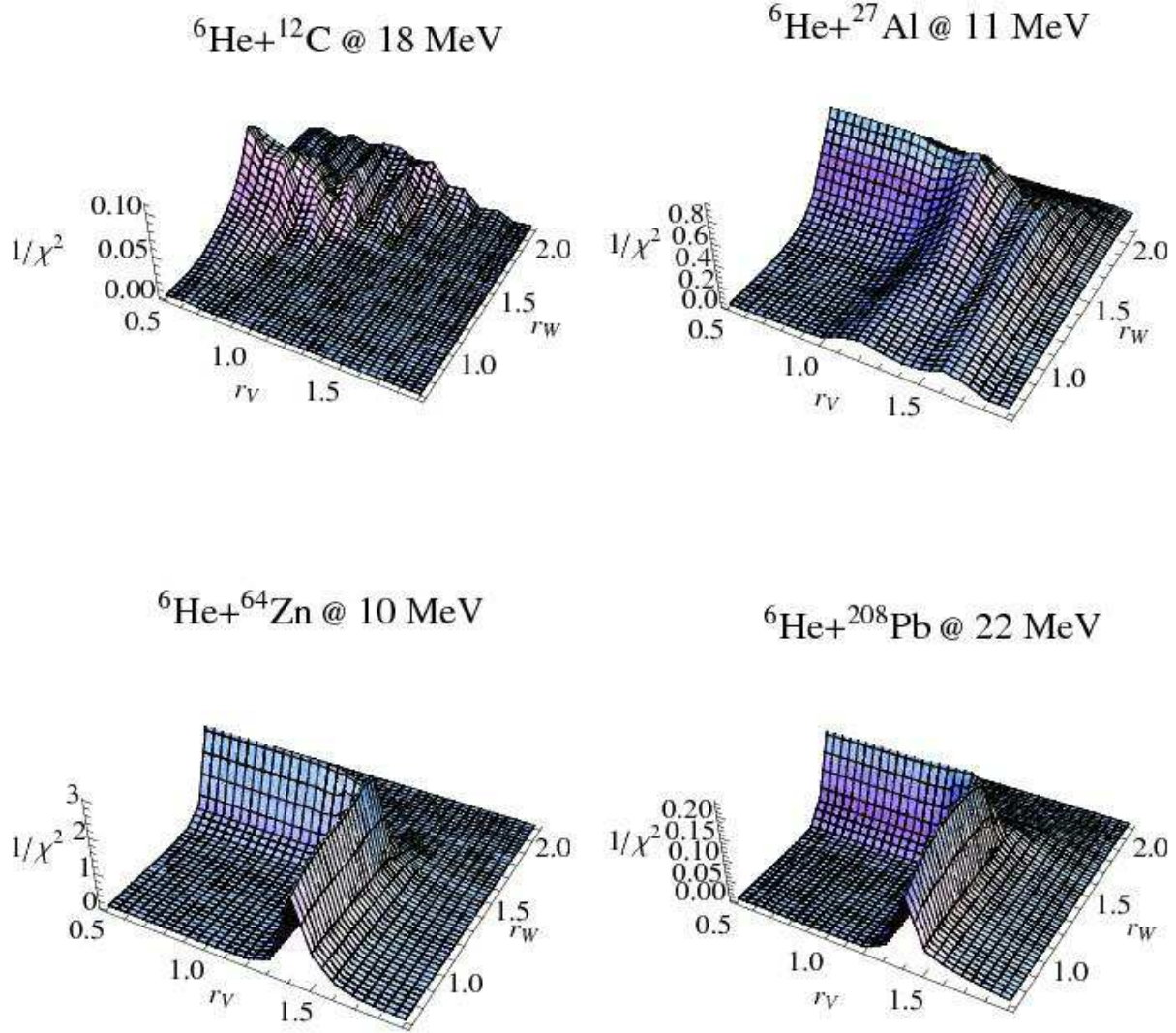}} \caption{(Color
online) Three-dimensional plots of the optical model parameters
$r_V$ , $r_W$ versus $1/\chi^2$, where $\chi^2$ has the usual
definition and measures the quality of the fit.} \label{fig1a}
\end{figure}

By taking free parameters of the depth of the real and imaginary
potential, we have investigated their radii for each part, which
give the best fit for the elastic scattering cross section data. In
order to perform this, we have made a $\chi^{2}$  search. The radii
of real ($r_{V}$) and imaginary potentials ($r_{W}$) have been
varied on a grid, respectively from 0.5 to 2.0 fm, with steps of 0.1
fm in order to obtain the best fit to the data \cite{kucukNPA}.The
results of this systematic search are shown in Figure~\ref{fig1a}
which is a three-dimensional plot of the $r_{V}$, $r_{W}$ and
1/$\chi^{2}$, where $\chi^{2}$ has the usual definition and measures
the quality of the fit. In Figure~\ref{fig1a}, the best fit
parameters, producing oscillating cross-sections with reasonable
phase and period, correspond to low $\chi^{2}$ values and peaks in
the 1/$\chi^{2}$ surface. For the four different reactions, the
figures present discrete peaks (or hills) for correlated $r_{V}$ and
$r_{W}$ values, which are best fit real and imaginary potential
families and indicate that the $r_{V}$ or $r_{W}$ parameters cannot
be varied continuously and still find equally satisfying fits. For
the radius of real part ($r_{V}$), the lowest $\chi^{2}$ values are
generally obtained around 0.9 fm and for the radius of imaginary
part ($r_{W}$), it is around 1.50 fm. The diffusion parameters have
also been fixed $a_{V}$=$a_{W}$=0.7~fm for both parts of the
potential.

Having obtained the best fit for all data, we have investigated the
change of the depth of the real and imaginary parts and we have
derived Equations \ref{real} and \ref{imag} for the variation of the
depth of the real and imaginary parts of the nuclear potential.
Equations depend on the incident energy of the projectile ($^{6}$He)
with the charge number ($Z$) and mass number($A$) of the target.

\begin{equation}
V_{0} = 110.1+2.1 \,\frac{Z_{T}}{A_{T}^{1/3}}+0.65\,E \label{real}
\end{equation}

\begin{equation}
W_{0} = 6.0+0.48 \,\frac{Z_{T}}{A_{T}^{1/3}}-0.15\,E \label{imag}
\end{equation}
where $E$ is the laboratory energy of the $^{6}$He and $Z_{T}$ and
$A_{T}$ are the charge and mass numbers of the target nuclei.

\begin{figure}
\epsfxsize 12.0cm \centerline{\epsfbox{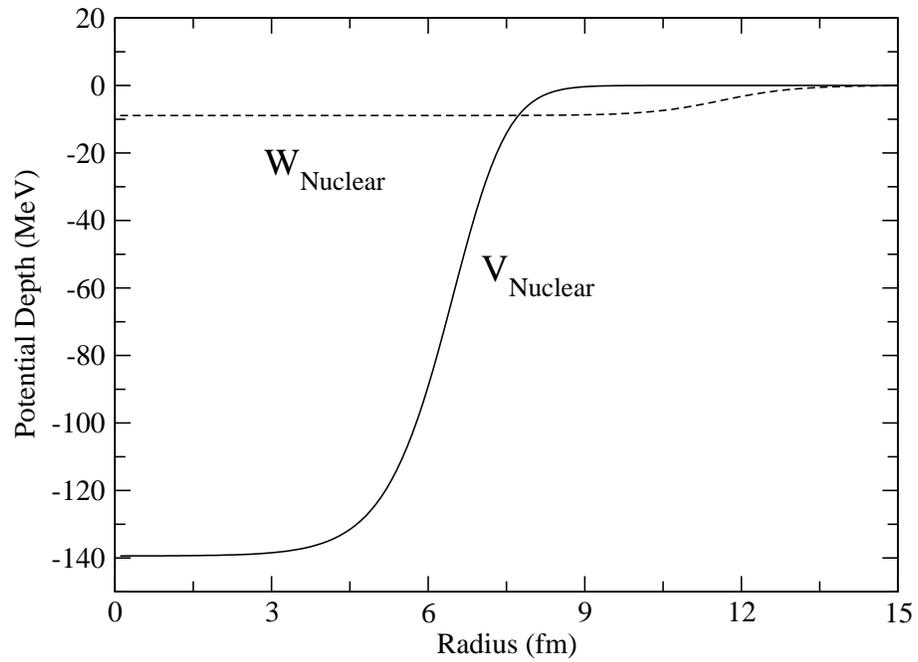}} \vskip+0.0cm
\caption{The real (solid line) and imaginary parts (dashed line) of
the nuclear potential at $E_{Lab}$=18 MeV for $^{6}$He + $^{208}$Pb.
} \label{fig1}
\end{figure}

For $^{6}$He+$^{208}$Pb system, the real and imaginary potentials
are shown in Figure~\ref{fig1} for $E_{Lab}$= 18.0 MeV. The sum of
the nuclear, Coulomb and the centrifugal potentials is also shown in
the same figure for the orbital angular momentum quantum numbers,
$l=0$ to $50$. The superposition of the attractive and repulsive
potentials results in the formation of a potential pocket, which the
width and depth of the pocket depend on the orbital angular
momentum. It is well known that this pocket is very important for
the interference of the barrier and internal waves, which produces
the pronounced structure in the cross-section~\cite{Bri77,BozOsi}.
We perceive from Figure~\ref{fig1} that the real part is located
inside the imaginary one, which shows that the long range absorption
is needed to explain the interaction of  $^{6}$He.
\section{Results} \label{results}
We have analyzed the elastic scattering of the $^{6}$He from target
nuclei of the $^{12}$C, $^{27}$Al,  $^{58}$Ni, $^{64}$Zn, $^{65}$Cu,
$^{197}$Au, $^{208}$Pb and $^{209}$Bi for a wide energy range below
50 MeV by using the derived new optical potential set given by Eqs.
\ref{real} and \ref{imag} within the framework of the optical model.

\begin{figure}
\epsfxsize 10.0cm \centerline{\epsfbox{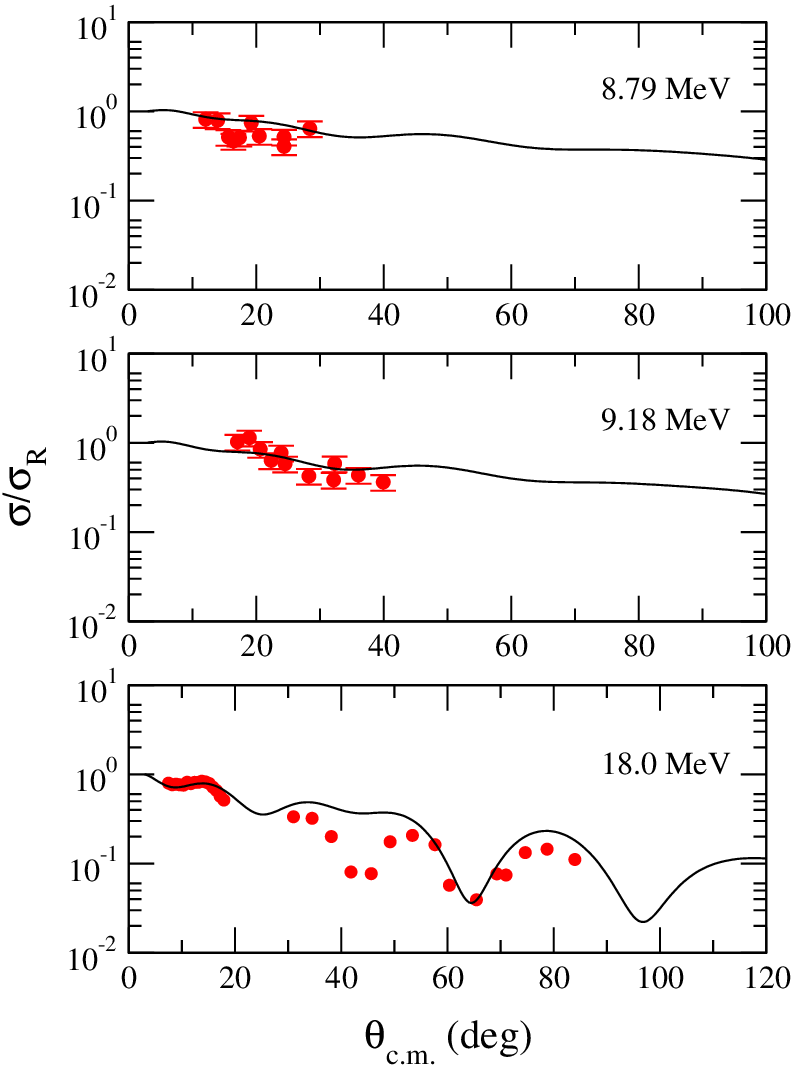}} \caption{ (Color
online) Elastic scattering angular distributions for $^{6}$He +
$^{12}$C. The solid lines show OM calculation results while the
circles show the experimental data. The experimental data have been
taken from Ref. \cite{Smith} and \cite{Milin}.\label{fig2}}
\end{figure}

First system we have considered is the $^{6}$He+$^{12}$C elastic
scattering,  an example of the light-heavy target, and we have
analyzed this system at 8.79, 9.18 and 18.0 MeV energies in the
laboratory system. The experimental data for 8.79 and 9.18 have been
measured by Smith \textit{et al.} \cite{Smith} and have been
analyzed by using the potential parameters of $^{4}$He, $^{6}$Li and
$^{7}$Li. In their work, the angular distribution of $^{6}$He has
been well produced by using $^{6}$Li and $^{7}$Li optical potential
parameters while $^{4}$He parameters have not produced the data
well. In our study, the elastic scattering data at these energies as
well as the data at 18.0 MeV measured by Milin \textit{et al.}
\cite{Milin} have been analyzed by using the new potential and a
good agreement has been obtained for all energies as presented in
Figure~\ref{fig2}. When the theoretical results are compared with
experimental data, we have obtained small $\chi^2/N$ values as it is
seen in Table~\ref{table1}. In the same table, we have also
presented the prediction of the new potential parameters for the
reaction cross section. The values are comparable with more
sophisticated CDCC or similar approaches.

\begin{figure}
\epsfxsize 8.0cm \centerline{\epsfbox{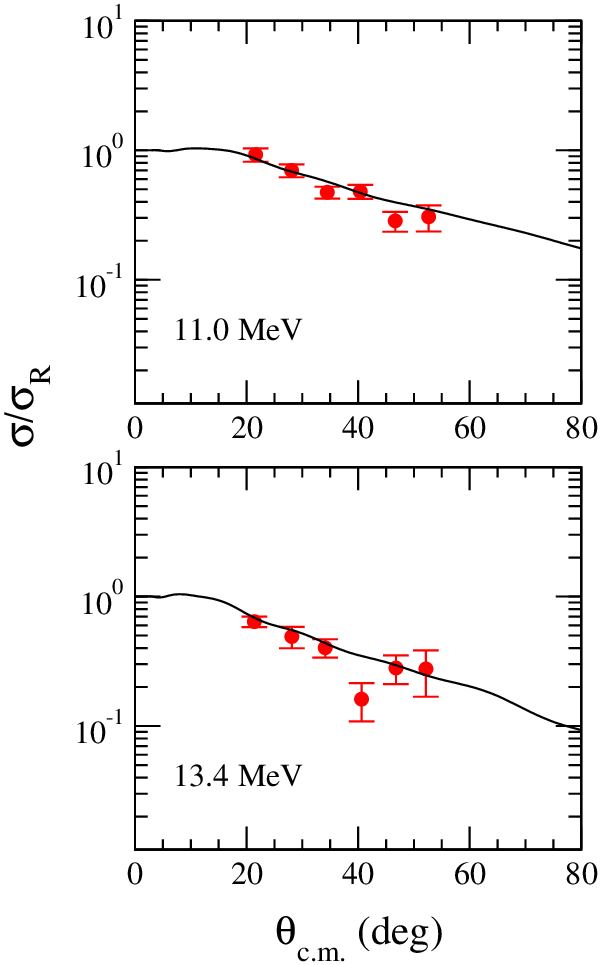}} \caption{(Color
online) Elastic scattering angular distributions (ratio to
Rutherford cross section) for $^{6}$He + $^{27}$Al. The solid lines
show OM calculation results while the circles show the experimental
data. The experimental data have been taken from Ref.
\cite{Benjamim}. \label{fig3}}
\end{figure}

Another studied system is the $^{6}$He+$^{27}$Al reaction. Elastic
scattering data of this system has been measured at energies 9.5,
11.0, 12.0, 13.4 MeV by using the RIBRAS (Radioactive Ion Beams in
Brazil) facilities by Benjamim \textit{et al.} \cite{Benjamim}.They
have also analyzed the measured data theoretically by using the Sao
Paulo Potential (SPP) and they have also deduced the reaction cross
section from the optical model fits. They have predicted the
reaction cross sections for these energies as 1110, 1257, 1300, 1327
mb, respectively. In comparing  our results with these values, we
see a difference of about 200 mb between the microscopic and our
phenomenological potentials. The difference is due to the shape of
the imaginary potential. The theoretical results of our potential
for $^{6}$He+$^{27}$Al elastic scattering and extracted reaction
cross sections for each energy are given in Table~\ref{table1} and
Figure~\ref{fig3}.

\begin{figure}
\epsfxsize 12.0cm \centerline{\epsfbox{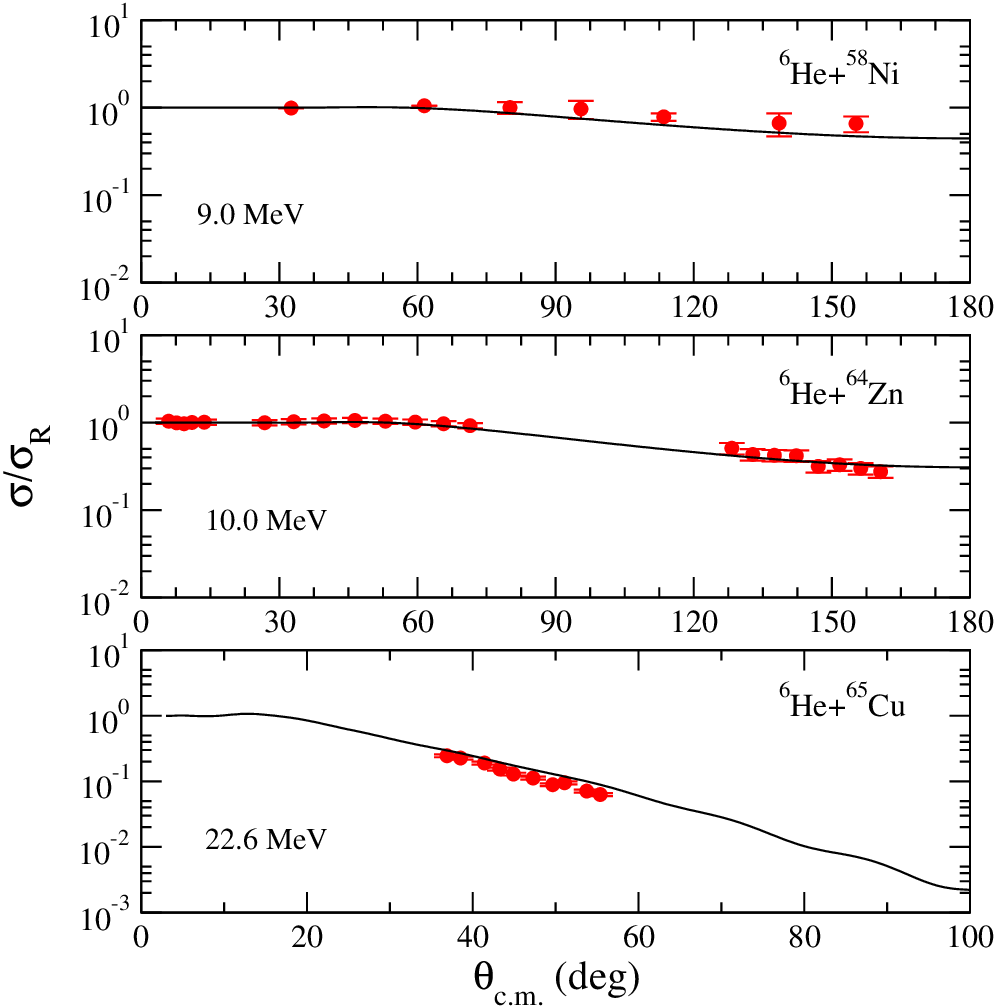}} \caption{(Color
online) Elastic scattering angular distributions (ratio to
Rutherford cross section) for $^{6}$He + $^{58}$Ni, $^{64}$Zn,
$^{65}$Cu. The solid lines show OM calculation results while the
circles show the experimental data. The experimental data have been
taken from Ref. \cite{Gasgues,Warner,Pietro} and \cite{Navin}.
\label{fig4}}
\end{figure}

For the medium mass target, $^{58}$Ni, $^{64}$Zn and $^{65}$Cu have
been analyzed by using the optical potential parameters obtained
from potential formula (Eqs. \ref{real} and \ref{imag}). These
systems have been studied around the Coulomb barrier and the elastic
scattering cross section have been measured by Refs.
~\cite{Gasgues,Pietro,Navin}. For $^{6}$He+$^{64}$Zn system, the
reaction cross section has been deduced as 380$\mp$60 mb for 10.0
MeV and 1450$\mp$130 mb for 13.6 MeV  using the phenomenological
potential set by Ref. \cite{Pietro}. These values are comparable
with  our results with a difference of around 10$\%$. The results of
our potential for the elastic scattering of these reactions and the
reaction cross section values for each energy are given in Table
\ref{table1} and Figure~\ref{fig4}.

\begin{figure}
\epsfxsize 12.0cm \centerline{\epsfbox{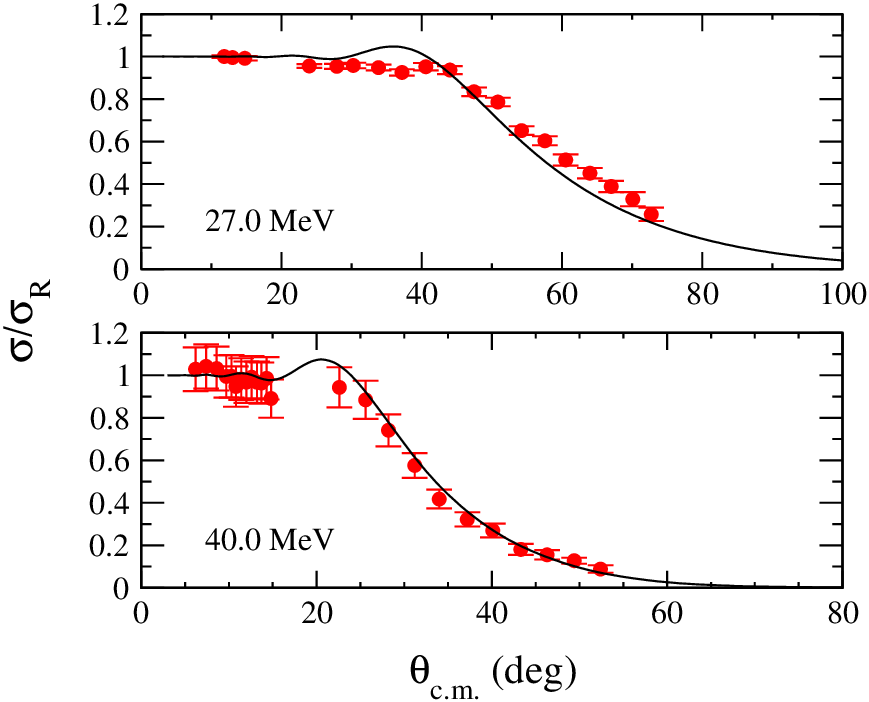}} \caption{(Color
online) Elastic scattering angular distribution (ratio to Rutherford
cross section) for $^{6}$He + $^{197}$Au. The solid lines show OM
calculation results while the circles show the experimental data.
The experimental data have been taken from Ref. \cite{Kakuee2006}
and \cite{Raab1}. \label{fig5}}
\end{figure}

\begin{figure}
\epsfxsize 12.0cm \centerline{\epsfbox{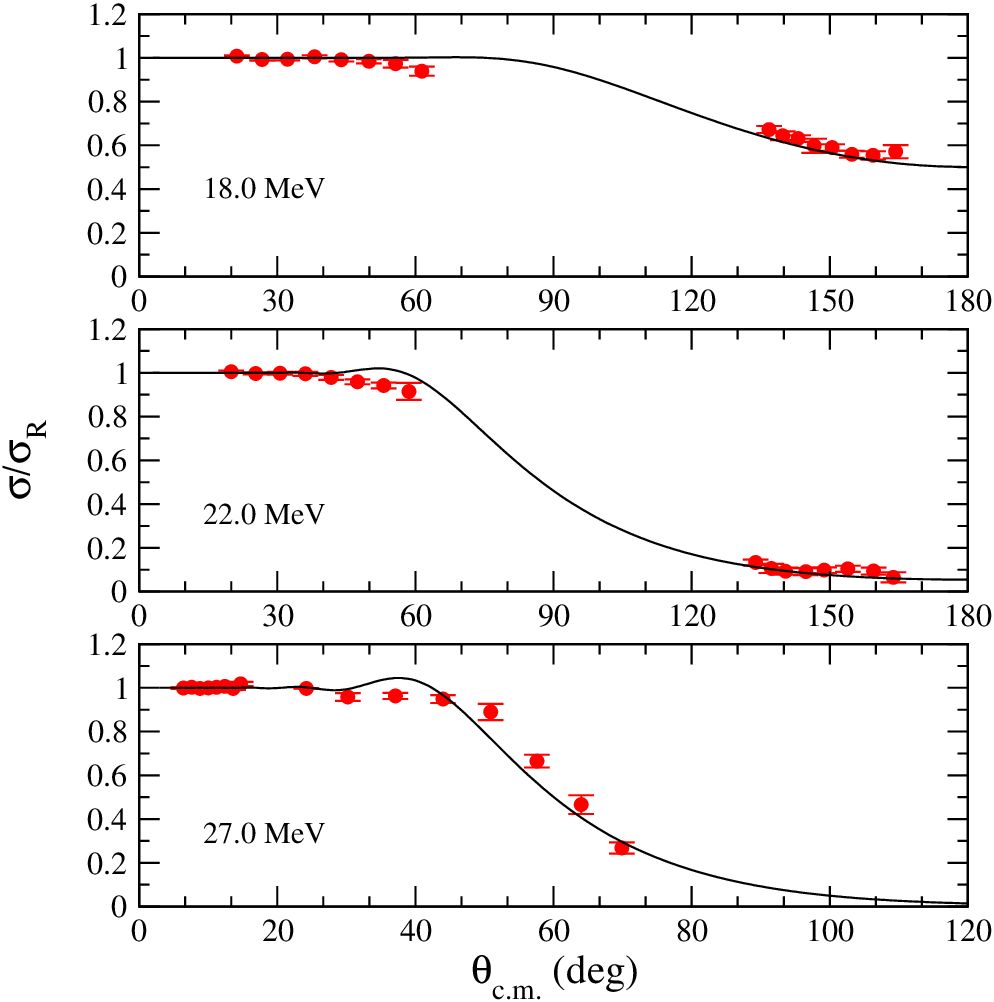}} \caption{(Color
online) Elastic scattering angular distribution (ratio to Rutherford
cross section) for $^{6}$He + $^{208}$Pb. The solid lines show OM
calculation results while the circles show the experimental data.
The experimental data have been taken from Ref.
\cite{Sanchez-Benitez2005} and \cite{Kakuee2003}. \label{fig6}}
\end{figure}

\begin{figure}
\epsfxsize 12.0cm \centerline{\epsfbox{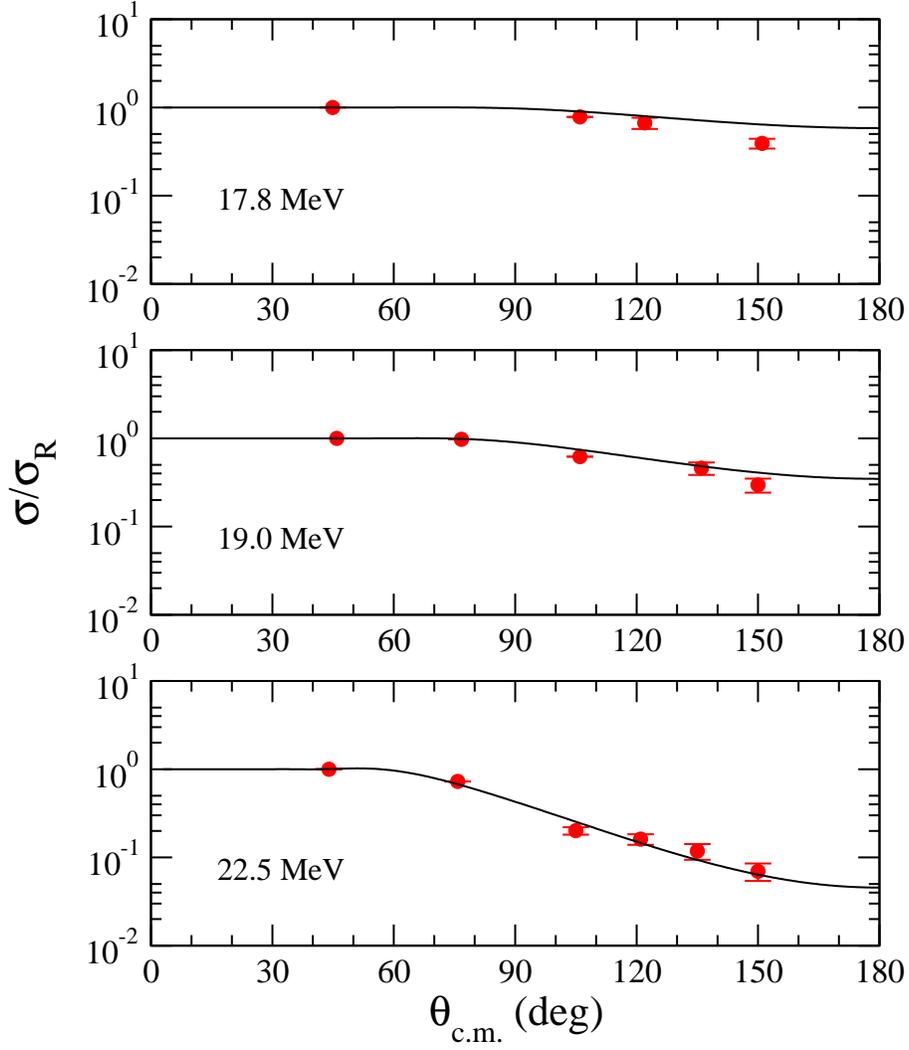}} \caption{(Color
online) Elastic scattering angular distribution (ratio to Rutherford
cross section) for $^{6}$He + $^{209}$Bi. The solid lines show OM
calculation results while the circles show the experimental data.
The experimental data have been taken from Ref. \cite{Agu1} and
\cite{Agu2}.\label{fig7}}
\end{figure}

We have also studied the elastic scattering of $^{6}$He from heavy
targets such as $^{197}$Au, $^{208}$Pb and $^{209}$Bi. For these
systems, the elastic scattering angular distributions have been
measured at energies near the Coulomb barrier generally. Kakuee
\textit{et al.} \cite{Kakuee2006} have measured the elastic
scattering cross section of $^{6}$He+$^{197}$Au and
$^{6}$He+$^{208}$Pb at 27.0 MeV and they have analyzed the data by
using the optical model. In their calculations, they have used the
parameters that fit the $^{6}$Li systems by both taking into account
and ignoring the dipole polarizability. However, since this
potential set was not adequate to fit the data, they have modified
the potential by changing the depth and diffuseness of the real
potential. They have shown that large imaginary diffuseness
parameters are required to fit the experimental data and they have
presented these results as an evidence of the long-range absorbtion
mechanism. In this work, they have found the reaction cross section
around 1900 mb at 27 MeV, which is estimated due to Coulomb
break-up. In comparing this work with our calculations, we observe
the same results. Our potential produces the reaction cross section
as 1925 mb and 1892 mb for $^{197}$Au and $^{208}$Pb nuclei at 27
MeV energy respectively. It also gives good results to elastic
scattering with very small $\chi^2/N$ values as shown in Figure
\ref{fig5}, Figure~\ref{fig6} and Table~\ref{table1}.  Beside these
works, the elastic scattering cross sections of $^{6}$He+$^{208}$Pb
and $^{6}$He+$^{209}$Bi for  eleven different energies conducted in
literature \cite{Sanchez,Agu1,Agu2} have been studied and excellent
agreement have been obtained for experimental data. The results for
some energies are given in Figures~\ref{fig6} and  ~\ref{fig7}.

\begin{figure}
\epsfxsize 16.0cm \centerline{\epsfbox{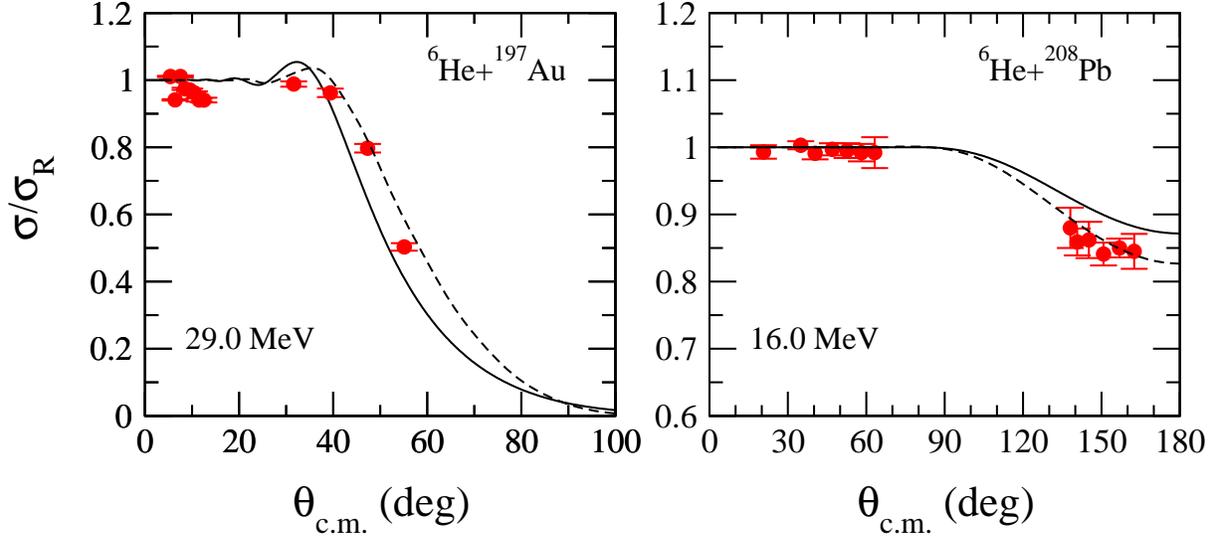}} \caption{(Color
online) Elastic scattering angular distribution (ratio to Rutherford
cross section) for $^{6}$He + $^{197}$Au and $^{6}$He + $^{208}$Pb.
The solid lines show OM calculation results with the imaginary
potential of Eq. \ref{imag} while the dashed lines show a decrease
(left panel) and an increase (right panel) of 5 MeV from the value
of Eq. \ref{imag}. \label{fig8}}
\end{figure}

We should point out that although the new potential set provides a
consistent agreement for many systems, the potential parameters need
a small modification to fit the experimental data at some energies
for heavy targets. As seen in Figures \ref{fig8}, while the new
potential predict the behavior of the cross section, it can not fit
the data exactly: It is over/under-estimate the experimental data at
particular cases. However, a change of the depth of the imaginary
potential such as $\pm$5 MeV is sufficient to fit the data.

In Table~\ref{table1}, for all reactions we have studied in this
paper, we have presented the $\chi^2/N$ values, the reaction cross
section values and the volume integrals produced by new potential
set. For the theoretical calculations, the code Fresco \cite{Fresco}
has been used.

\section{Conclusion} \label{conc}

We have presented a new potential set by deriving a formula for the
depth of the real and imaginary parts of the optical potential for
$^{6}$He elastic scattering at low energies. We should point out
that we do not aim to obtain the best fits for the experimental
data. Rather, we attempt to derive a global potential set that
produces the behavior of the experimental data reasonably well. In
this sense, we have analyzed almost all experimental data conducted
over a wide energy range in the literature by using this potential
set to show the validity of the potential in explaining the elastic
scattering data and reasonable agreement has been obtained for all
data with reasonable $\chi^2/N$ values.

As it may be seen from our results obtained by using the new
potential parameters, one can easily use this potential set instead
of using the improved parameters of the most similar nuclei such as
$^{4}$He, $^{6}$Li and $^{7}$Li as it is most commonly done. This
global potential can also be extended to describe the scattering
observables of other halo type nuclei which is important in
providing information regarding their interaction mechanism.
\section*{Acknowledgments}
This work has been supported by the Turkish Science and Research
Council (T\"{U}B\.{I}TAK) with Grant No:107T824, the Turkish Academy
of Sciences (T\"{U}BA-GEB\.{I}P) and Akdeniz University Scientific
Research Projects Unit.

\newpage

\begin{table}

\begin{center}
\begin{tabular}{|cccccc|cccccc|}
    \hline
& E & $\chi^2/N$ & $\sigma_{_R}$&  $J_{V}$ & $J_{W} $&   & E & $\chi^2/N$ & $\sigma_{_R}  $ &$J_{V}$ &  $J_{W} $ \\
& $MeV$&  & $mb$&  $MeVfm^3$ & $MeVfm^3 $&   & $MeV$ &  & $mb$ &$MeVfm^3$ &  $MeVfm^3 $ \\

    \hline
$\quad ^{12}C$ & 8.79  &4.76&1241. &256.24 &90.91 $\quad$& $\quad^{27}Al$& 9.5  & 1.14 &1237. &193.22 &70.87$\quad$ \\
& 9.18  &2.52 &1262. &256.79 & 90.14$\quad$&  & 11.0 & 1.61 &1380. &194.71     & 68.42$\quad$ \\
& 18.0  &26.30 &1453. &268.90 &69.75$\quad$&  & 12.0 & 4.25&1453. &195.71      &66.82$\quad$\\
             &       & &            &      && & 13.4 & 2.29 &1531.  &197.12     & 64.58$\quad$\\
\hline
 $\quad^{64}Zn$& 10.0  &0.40&537.5 &155.73 &62.35$\quad$& $\quad^{65}Cu$ &  22.6 &36.27&2012.&163.95 &46.53$\quad$\\

\hline

$\quad^{58}Ni$ &  9.0 & 1.63&383.1 &158.63 &64.73 $\quad$& $\quad^{197}Au$& 27.0 & 13.12&1925. &141.2 &46.70$\quad$\\
                   &       & $\quad$&     &                    &&  & 40.0 &0.51 &2843. &148.88& 35.94$\quad$\\
    \hline
    $\quad^{208}Pb$& 14.0  &0.75&3.935 &132.69 &57.39$\quad$&$\quad^{209}Bi$ & 14.7 & 19.87&9.190&133.25 &67.10$\quad$\\
                   & 16.0  &3.11&59.71 &133.85 &55.76$\quad$&              & 16.3& 14.93&64.01 &134.15 & 55.85$\quad$\\
                   & 18.0  &3.23&302.3 &135.02 &54.12$\quad$&             &17.8& 11.40&235.7 &135.05 &54.60$\quad$\\
                   & 22.0  &5.05&1118. &137.34 &50.85$\quad$&             & 19.0& 2.27&454.6 &135.70  & 53.62$\quad$\\
                   & 27.0  & 5.70&1892. &140.71 &46.82$\quad$&        & 22.5& 2.25&1168. &137.78 & 50.74$\quad$\\

 \hline
  \end{tabular}
  \end{center}
\caption{The reaction cross section, volume integrals and $\chi^2/N$
values obtained by using the experimental error bars.}
\label{table1}
\end{table}

\newpage


\end{document}